\documentclass[aps,amsmath,amssymb,floatfix,onecolumn, showpacs]{revtex4-1}
\usepackage{graphicx}
\usepackage{dcolumn}
\usepackage{bm}
\usepackage{color}
\usepackage{placeins}
\usepackage{array}
\usepackage{epstopdf}
\usepackage{subfigure}
\usepackage{natbib}
\usepackage{amssymb}
\usepackage{amsbsy}
\usepackage{citesort}

\begin{document}

\title{Effect of aspect-ratio on vortex distribution and heat transfer in rotating Rayleigh-B\'enard convection}
\author{Richard J.A.M. Stevens$^1$}
\author{Jim Overkamp$^2$}
\author{Detlef Lohse$^{1}$}
\author{Herman J.H. Clercx$^{2,3}$}
\affiliation{$^1$Department of Science and Technology and J.M. Burgers Center for Fluid Dynamics, University of Twente, P.O Box 217, 7500 AE Enschede, The Netherlands}
\affiliation{$^2$Department of Physics and J.M. Burgers Center for Fluid Dynamics, Eindhoven University of Technology, P.O. Box 513, 5600 MB Eindhoven, The Netherlands}
\affiliation{$^3$Department of Applied Mathematics, University of Twente, P.O Box 217, 7500 AE Enschede, The Netherlands}

\date{\today}

\begin{abstract}
Numerical and experimental data for the heat transfer as function of the Rossby number $Ro$ in turbulent rotating Rayleigh-B\'enard convection are presented for Prandtl number $Pr=4.38$ and Rayleigh number $Ra=2.91\times10^8$ up to $Ra=4.52\times10^9$. The aspect ratio $\Gamma \equiv D/L$, where $L$  is the height and $D$ the diameter of the cylindrical sample, is varied between $\Gamma=0.5$ and $\Gamma=2.0$. Without rotation, where the aspect ratio influences the global large scale circulation, we see a small aspect-ratio dependence in the Nusselt number for $Ra=2.91\times10^8$. However, for stronger rotation, i.e. $1/Ro \gg 1/Ro_c$, the heat transport becomes independent of the aspect-ratio. We interpret this finding as follows: In the rotating regime the heat is mainly transported by vertically-aligned vortices. Since the vertically-aligned vortices are local, the aspect ratio has a negligible effect on the heat transport in the rotating regime. Indeed, a detailed analysis of vortex statistics shows that the fraction of the horizontal area that is covered by vortices is independent of the aspect ratio when $1/Ro \gg 1/Ro_c$. In agreement with the results of Weiss \emph{et al.} (Phys. Rev. Lett., vol 105, 224501 (2010)) we find a vortex-depleted area close to the sidewall. Here, we in addition show that there is also an area with enhanced vortex concentration next to the vortex-depleted edge region and that the absolute widths of both regions are independent of the aspect ratio.
 \end{abstract}


\maketitle

\section{Introduction}

The classical system to study turbulent thermal convection in  confined space is the Rayleigh-B\'enard (RB) system, i.e., fluid between two parallel plates heated from below and cooled from above \cite{ahl09,loh10}. For given aspect ratio $\Gamma\equiv D/L$ ($D$ is the sample diameter and $L$ its height) and given geometry, its dynamics are determined by the Rayleigh number $Ra=\beta g\Delta L^3 /(\kappa \nu)$ and the Prandtl number $Pr=\nu/\kappa$. Here $\beta$ is the thermal expansion coefficient, $g$ the gravitational acceleration, $\Delta$ the temperature difference between the plates, and $\nu$ and $\kappa$ are the kinematic and thermal diffusivity, respectively. The case where the RB system is rotated around a vertical axis at an angular speed $\Omega$, i.e. rotating Rayleigh-B\'enard (RRB) convection, is interesting for industrial applications and problems in geology, oceanography, climatology, and astronomy. The rotation rate of the system is non-dimensionalized in the form of the Rossby number $Ro=\sqrt{\beta g \Delta/L}/(2\Omega)$, which represents  the ratio between buoyancy and the Coriolis force.

It is well know that the boundary layer behavior plays an important roll in the heat transfer properties of a RB system \cite{ahl09,gro00,gro01,gro02,gro04}. However, several studies have shown that also the general flow structure in the system can influence the overall heat transfer. For non-rotating RB Sun, Xi and Xia \cite{sun05a} were the first to demonstrate that the global transport in any turbulent flow system can depend on internal flow structures. Later Xi and Xia \cite{xi08} quantified the difference in Nusselt between the Single Roll State (SRS) and Double Roll State (DRS). Recently, Weiss and Ahlers \cite{wei10b} found that the difference in the heat transport between the SRS and DRS decreases with increasing $Ra$. Subsequently, Van der Poel et al.\ \cite{poe11} found that in 2D RB convection a different flow structure can change the heat transfer up to $30\%$. An overview on the importance of the internal flow structure on the heat transfer is given by Xia \cite{Xia11}. The mechanism of the effect of the bulk flow on the heat transfer is as follows: More efficient bulk flow transports hot (cold) plumes to the top (bottom), thus creating a larger mean temperature gradient there, and thus a larger $Nu$. 
In this paper we will see that also in rotating RB convection the internal flow structures influence the overall heat transport in the system \cite{bou90,zho09b,ste09,wei10,kun10,zho10c,kun11}.

It has been shown by several authors that three different regimes can be identified in RRB convection \cite{bou90,zho09b,ste09,wei10,kun10,zho10c,kun11}. As function of increasing rotation rate one first finds a regime without any  heat transport enhancement at all in which the  large scale circulation (LSC) is still present (regime I). Zhong $\&$ Ahlers \cite{zho10c} showed that -- though the Nusselt number is unchanged in this regime -- nonetheless  various properties of the LSC do change with increasing rotation in this regime. Here we mention the increase in the temperature amplitude of the LSC, the LSC precession (also observed by Hart et al.\ \cite{har02} and Kunnen et al.\ \cite{kun08b}), the decrease of the temperature gradient along the sidewall, and the increased frequency of cessations. The start of regime II (moderate rotation) is indicated by the onset of heat transport enhancement due to Ekman pumping as is discussed by Zhong et al.\ \cite{zho09b} and Weiss et al.\ \cite{wei10,wei11}. When the rotation rate is increased in regime II the heat transfer increases further until one arrives at regime III (strong rotation), where the heat transfer starts to decrease. This decrease of the heat transfer in regime III is due to the suppression of the vertical velocity fluctuations \cite{zho09b,zho10c,ste10a}.

\begin{figure}
\centering
\subfigure{\includegraphics[width=0.49\textwidth]{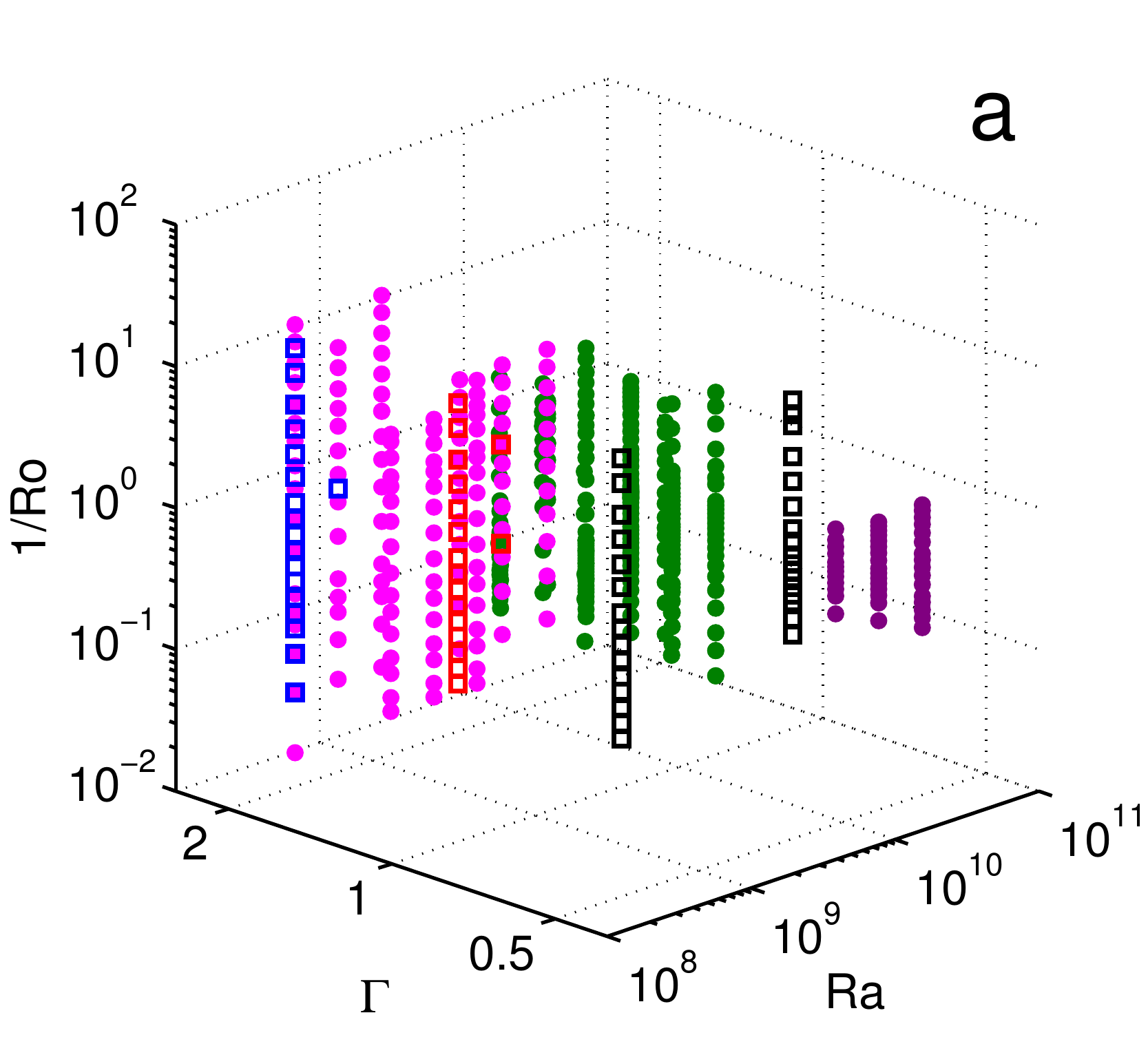}}
\subfigure{\includegraphics[width=0.49\textwidth]{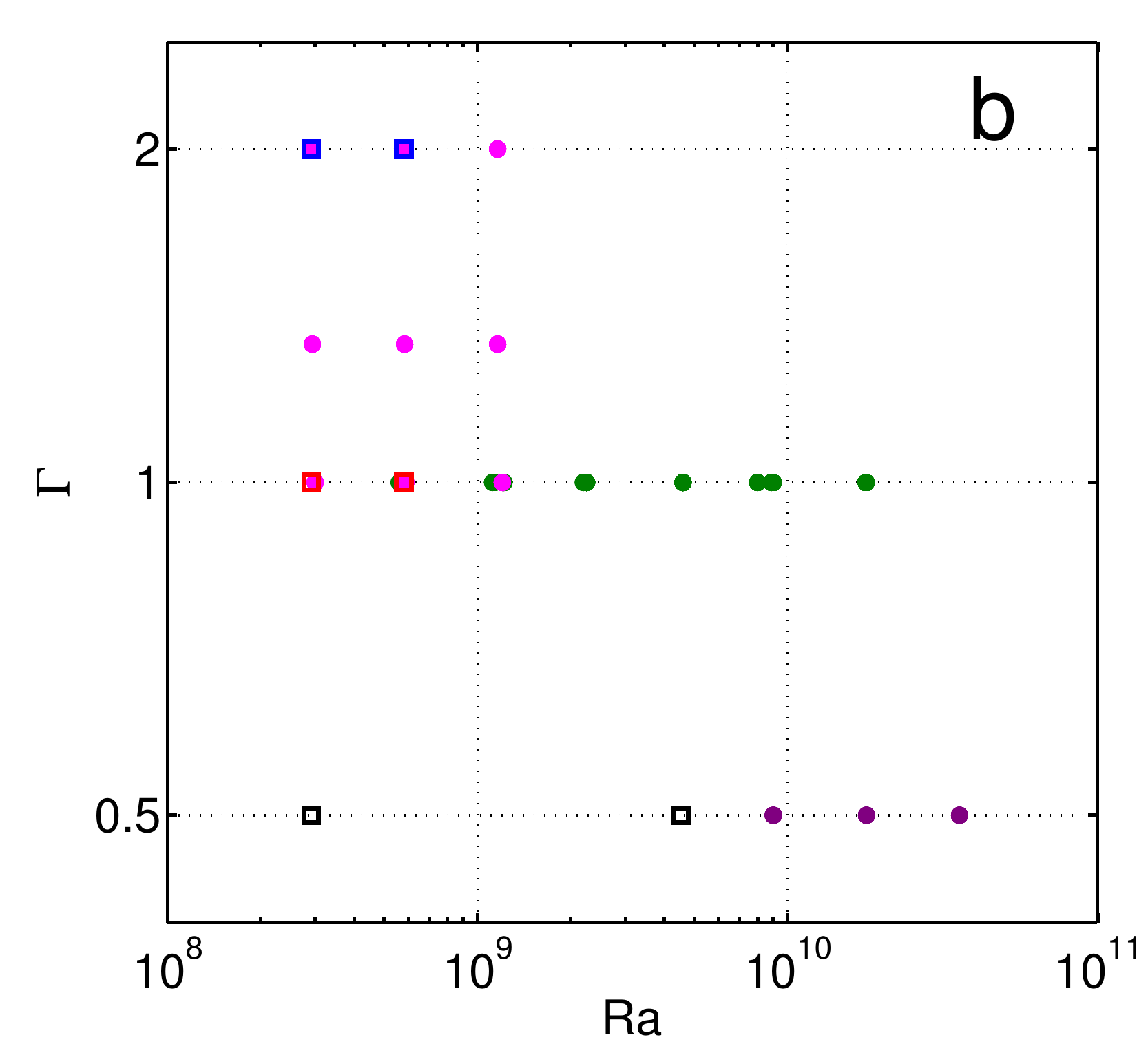}}
\subfigure{\includegraphics[width=0.49\textwidth]{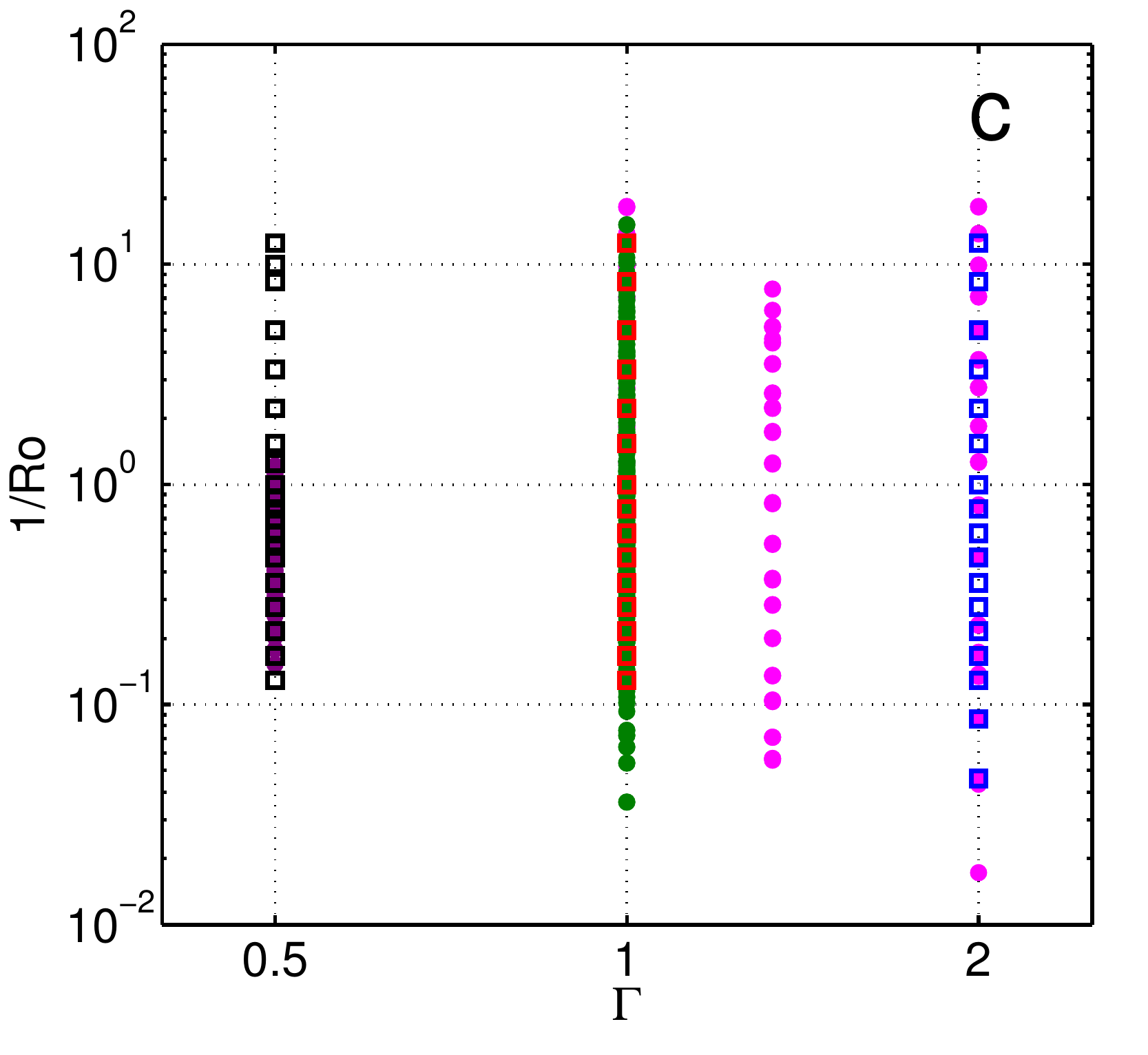}}
\subfigure{\includegraphics[width=0.49\textwidth]{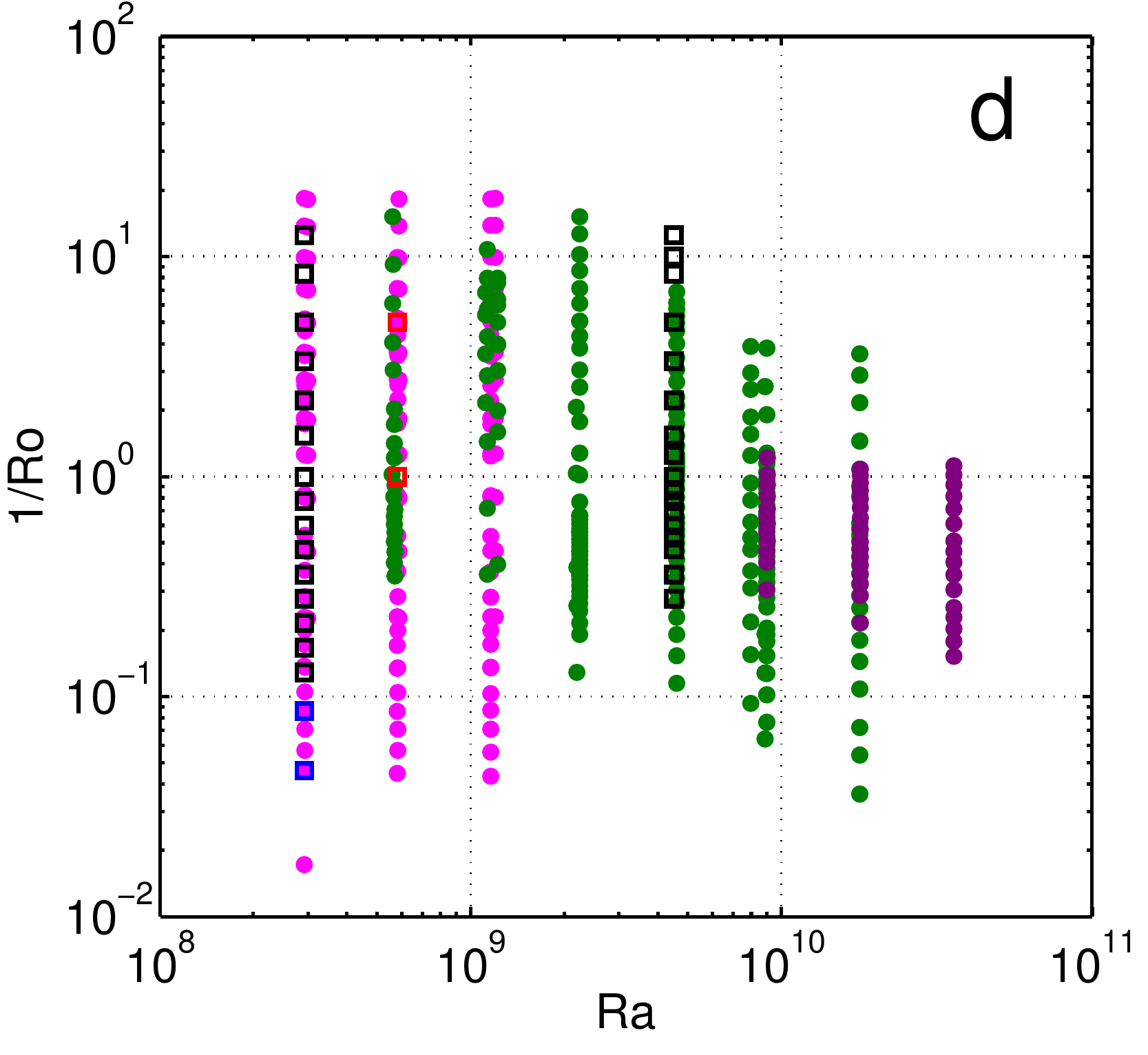}}
\subfigure{\includegraphics[width=0.95\textwidth]{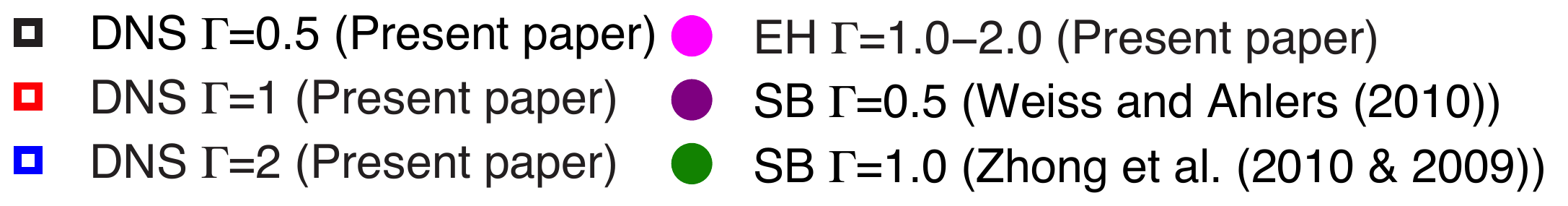}}
\caption{
Phase diagram in $Ra-Ro-\Gamma$ space for RRB convection with $Pr=4.38$. The data points indicate where $Nu$ has been experimentally measured or numerically calculated in a cylindrical sample with no-slip boundary conditions. The DNS data and the Eindhoven (EH) experiments are from this study. The experimental results from Santa Barbara (SB) are from Weiss et al.\ 
\cite{wei10} and Zhong $\&$ Ahlers \cite{zho10c} for the $\Gamma=0.5$ and $\Gamma=1.0$, respectively.
The EH experimental data focus on $\Gamma > 1$, but for one case we also give
the result for $\Gamma = 1$ for benchmarking with the earlier SB data. 
 a) A three dimensional view on the phase space (see also the supplementary material),
 b) Projection on $\Gamma-Ra$ phase space,
 c) Projection on $1/Ro-\Gamma$ phase space and,
 d) Projection on $1/Ro-Ra$ phase space.
 We also refer to  figure 1 of Stevens et al.\ \cite{ste10b} 
for a full representation of 
the $Ra-Pr-Ro$ phase space of  RRB convection.
}
\label{fig:Phase space_gamma}
\end{figure}

Several experimental
\cite{ros69,liu97,liu09,kin09,zho09b,ste09,zho10c}
and numerical
\cite{jul96,spr06,ore07,kun08b,kun10,kin09,sch09,zho09b,ste09,sch10,ste10a}
studies on RRB convection have shown that in regime II the heat transport with respect to the non-rotating case increases due to rotation. A detailed overview of the parameter ranges covered in the different experiments can be found in the $Ra-Pr-Ro$ phase diagram shown in figure 1 of Stevens et al.\ \cite{ste10b} and the $Ra-Ro-\Gamma$ phase diagram shown in figure \ref{fig:Phase space_gamma}.
The heat transport enhancement in regime II is caused by Ekman pumping
\cite{ros69,zho93,jul96,vor02,kun08b,kin09,zho09b,zho10c}.
Namely, whenever a plume is formed, the converging radial fluid motion at the base of the plume (in the Ekman boundary layer (BL)) starts to swirl cyclonically, resulting in the formation of vertical vortex tubes. The rising plume induces stretching of the vertical vortex tube and hence additional vorticity is created. This leads to enhanced suction of hot fluid out of the local Ekman layer and thus increased heat transport. Corresponding phenomena occur at the upper boundary. 

Zhong et al.\ \cite{zho09b} and Stevens et al.\ \cite{ste09,ste10a} used results from experiments and direct numerical simulations (DNS) in a $\Gamma=1$ sample to study the influence of $Ra$ and $Pr$ on the effect of Ekman pumping. It was found that at fixed $Ro$ the effect of Ekman pumping is largest, and thus the observed heat transport enhancement with respect to the non-rotating case highest, at an {\it intermediate} Prandtl number. At lower $Pr$ the effect of Ekman pumping is reduced as more hot fluid that enters the vortices at the base spreads out in the middle of the sample due to the large thermal diffusivity of the fluid. At higher $Pr$ the thermal BL becomes thinner with respect to the kinetic BL, where the base of the vortices is formed, and hence the temperature of the fluid that enters the vortices becomes lower. In addition, it was found that the effect of Ekman pumping is reduced for increasing $Ra$. This is because the turbulent viscosity increases with increasing $Ra$, which means that more heat spreads out in the middle of the sample. This explains why for  the high $Ra$-number of Niemela et al.\ \cite{nie10} no heat transfer enhancement was found in the rotating case. From this we can conclude that the internal flow structures influence the heat transfer in rotating RB as is also the case for non-rotating RB convection \cite{sun05a,xi08,wei10b,poe11}.

Recently, Weiss et al.\ \cite{wei10,wei11} showed that the rotation rate at which the onset of heat transport enhancement sets in ($1/Ro_c$) increases with decreasing aspect ratio due to finite size effect. This means that the aspect ratio is an important parameter in RRB convection. We report in this paper on a systematic study of the influence of the aspect ratio on heat transfer (enhancement) for moderate and strong rotation rates, i.e., for  $1/Ro \gg 1/Ro_c$. We start this paper with a description of the experimental and numerical procedures that have been followed. Where the experimental and simulation results overlap an excellent agreement is found. Based on the numerical data we will show that  for $1/Ro \gg 1/Ro_c$ the Nusselt number is independent of the aspect ratio, while there are some visible differences for the non-rotating case. The reason for this is that in the non-rotating case there is a global flow organization, which can be influenced by the aspect ratio. In the rotating regime vertically aligned vortices, in which most of the heat transport takes place \cite{por08,gro10}, form the dominant feature of the flow. As this is a local effect the heat transport in this regime does not depend on the aspect ratio. In the last part of the paper we will analyze the vortex statistics for the different aspect ratios to support that the vortices are indeed a local phenomenon.

\section{Experimental procedure} \label{Sec2}
Various heat transport measurements on RRB convection were performed in the group of Guenter Ahlers in Santa Barbara. These measurements were done in an aspect ratio $\Gamma=1$ sample and cover the $Ra$ number range $3 \times 10^8 \lesssim Ra \lesssim 2 \times 10^{10}$, the $Pr$ number range $3.0  \lesssim Pr \lesssim 6.4$, and the $1/Ro$ number range $0 \lesssim 1/Ro \lesssim 20$. These data points are shown in the $Ra-Ro-\Gamma$ phase diagram for RRB shown in figure \ref{fig:Phase space_gamma}. The experimental procedure that has been used in these experiments is described in detail by Zhong $\&$ Ahlers \cite{zho10c}. All the Nusselt number measurements of Zhong $\&$ Ahlers \cite{zho10c} were documented in the accompanying supplementary material of that paper, and we use those data in our figures here. Recently, Weiss $\&$ Ahlers \cite{wei11} did similar measurements in an aspect ratio $\Gamma = 0.5$ sample. As these measurement data are not yet available in the open literature we do not include a comparison with their data.

Zhong $\&$ Ahlers \cite{zho10c} restricted themselves to aspect ratio $\Gamma=1$. Here  we present new heat transport measurements from the Eindhoven RB setup \cite{kun11}, which is based on the Santa Barbara design \cite{bro05,zho10c}, for aspect ratio $\Gamma=4/3$ and $\Gamma=2.0$. We also present one $\Gamma=1.0$ measurement for one particular relatively small $Ra$ number from the Eindhoven setup, (i) because this particular $Ra$ number was not available in the Santa Barbara (SB) data set of Zhong $\&$ Ahlers \cite{zho10c} and we wanted to compare the results with those for $\Gamma = 2$ and $\Gamma=4/3$ at the same $Ra$, (ii) because we want to compare with the simulation results which are restricted to small $Ra$, and (iii) because we wanted to benchmark the results of our new setup against those of the SB setup. 

The Eindhoven RB convection sample has a diameter $D$ of 250 mm and due to the modular design of the setup measurements for different aspect ratios ($\Gamma=1.0-2.0$) can be performed by replacing the sidewall. All measurements are performed with a mean temperature of $40^\circ C$ ($Pr=4.38$).  For any given data point, measurements over typically the first four hours were discarded to avoid transients, and data taken over an additional period of at least another eight hours were averaged to get the average plate temperatures and the required power. Details about the setup and the experimental procedure, which is closely based on the Santa Barbara one \cite{bro05,zho10c}, can be found in Kunnen et al.\ \cite{kun11}.

Presently, the Eindhoven experiments can only be performed with one type of plate material (copper) and not with more as done in Santa Barbara. For this reason we can not apply the plate corrections to obtain the absolute Nusselt number without making some assumptions that cannot be verified at the moment. Therefore for the experimental results we restrict ourself to relative Nusselt numbers $Nu(1/Ro)/ Nu(0)$.

\section{Numerical procedure} \label{Sec3}

\begin{table}
  \centering
  \caption{Simulations performed in this study. The columns from left to right indicate the following: the Rayleigh number $Ra$, the aspect ratio $\Gamma$, the inverse Rossby number $1/Ro$, the number of $Ro$ cases that is simulated ($n_{Ro}$), and the number of grid points in the azimuthal, radial, and axial directions ($N_{\theta} \times N_r \times N_z$) }
  \label{table1}
\begin{tabular}{|c|c|c|c|c|}
\hline
$Ra$ 					& $\Gamma $ 	& $1/Ro$ 		& $n_{Ro}$ 	& $N_{\theta} \times N_r \times N_z$\\
 \hline
  $2.91 \times 10^8$			&  $0.5$ 		& $0-12.5$	& $16$	   	& $  257    \times   97	\times 289$\\
  $4.52 \times 10^9$			&  $0.5$ 		& $0-12.5$	& $18$		& $  641    \times 161	\times 641$\\
  $2.91 \times 10^8$			&  $1.0$		& $0-12.5$	& $16$		& $  385    \times 193	\times 289$\\
  $5.80 \times 10^8$			&  $1.0$		& $0-5.0$		& $3$		& $  641    \times 193	\times 385$\\  
  $2.91 \times 10^8$			&  $2.0$		& $0-12.5$	& $18$		& $  769	\times  385	\times 289$\\    
  $5.80 \times 10^8$			&  $2.0$		& $0-5.0$		& $2$		& $1281	\times  385	\times 385$\\ 
    \hline
\end{tabular}
\end{table}

In the simulations the flow characteristics of RRB convection for $Ra=2.91\times 10^8 - 4.52\times10^9$, $Pr=4.38$, $0<1/Ro<12.5$, and $0.5<\Gamma<2.0$, see also table \ref{table1} and figure \ref{fig:Phase space_gamma}, are obtained from solving the three-dimensional Navier-Stokes equations within the Boussinesq approximation:

\begin{eqnarray}
 \frac{D\textbf{u}}{Dt} &=& - \nabla P + \left( \frac{Pr}{Ra} \right)^{1/2} \nabla^2 \textbf{u} + \theta \textbf{$\widehat{z}$}- \frac{1}{Ro} \widehat{z} \times \textbf{u}, \\
 \frac{D\theta}{Dt} &=& \frac{1}{(PrRa)^{1/2}}\nabla^2 \theta ,
\end{eqnarray}
with
 $\nabla \cdot \textbf{u} = 0$.
 Here \textbf{$\widehat{z}$} is the unit vector pointing in the opposite direction to gravity,
 $D/Dt = \partial_t + \textbf{u} \cdot \nabla $ the material derivative,
 $\textbf{u}$ the velocity vector (with no-slip boundary conditions
 at all walls), and $\theta$ the non-dimensional temperature, $0\leq \theta \leq 1$.
 Finally, $P$ is the reduced pressure (separated from its hydrostatic contribution, but containing the centripetal contributions): $P=p - r^2/(8Ro^2)$, with $r$ the distance to the rotation axis. The equations have been made non-dimensional by using, next to $L$ and $\Delta$, the free-fall velocity $U=\sqrt{\beta g \Delta L}$. A constant temperature boundary condition is used at the bottom and top plate and the side wall is adiabatic. Further details about the numerical procedure can be found in Refs. \cite{ver96,ver99,ver03}.
   
The resolutions used for the simulations are summarized in table \ref{table1}. These grids allow for a very good resolution of the small scales both inside the bulk of turbulence and in the BLs where the grid-point density has been enhanced. We checked this by calculating the convergence of the volume averaged kinetic $\epsilon_u$ and thermal $\epsilon_\theta$ dissipation rates as is proposed by Stevens et al.\ \cite{ste10}.  We find that these quantities always converge within a $5\%$ margin at most. A comparison with the results of Stevens et al.\ \cite{ste10} shows that this convergence rate is clearly sufficient to reliably calculate the heat transfer. As argued by Shishkina et al.\ \cite{shi10} it is especially important to properly resolve the BLs. According to equation (42) of that paper the minimal number of nodes that should be placed in the thermal BL is $N_{BL}=4.7$ ($N_{BL}=7.1$) when $Ra=2.91\times10^8$ ($Ra=4.52\times10^9$) and $Pr=4.38$. In the simulations at $Ra=2.91\times10^8$ ($Ra=4.52\times10^9$) we placed $N_{BL}=14$ ($N_{BL}=22$) points in the thermal BL, which is on the safe side. For the non-rotating case similar numbers are obtained for the grid point resolution in the kinetic BL. When rotation is applied the kinetic BL becomes thinner. This effect only becomes significant for the highest $1/Ro$ number cases considered here, and we emphasize that for all cases the number of points in the kinetic BL is above the criterion put forward by Shishkina et al.\ \cite{shi10}. Furthermore, it is also very important to make sure that the results are statistically converged. Again we use the methods introduced by Stevens et al.\ \cite{ste10} to check this. For all cases the convergence is in the order of $1\%$ and it is much better for most. For the simulations at $Ra=2.91\times10^8$ in the aspect ratios $\Gamma=0.5$, $\Gamma=1.0$, and $\Gamma=2.0$ the average statistical uncertainty is only about $0.5\%$. 

As a last check we compare the numerical results with experimental data. For this we use the data of the group of Guenter Ahlers, who did high precision heat transport measurements in $\Gamma=0.5$ \cite{wei10b}, $\Gamma=1.0$ \cite{fun05}, and $\Gamma=2.0$ \cite{fun05} samples. For all cases we find that the numerical results are within $1\%$ of these experimental data. For some cases the numerical data of the present simulations are even slightly below the experimental data. This is reassuring with respect to numerical resolution issues, as normally the heat transport in underresolved simulation is {\it larger} than the actual heat transport, see Stevens et al.\ \cite{ste10}. 

In the simulations we partially neglect centrifugal forces, namely the density dependence of the centripetal forces, which in the Boussinesq equations show up as  $-2 Fr ~ r\theta \widehat{r}$, with the radial unit vector $\hat{r}$ \cite{hom69}. Several authors \cite{kun06,kun08b,zho09b} have shown that this is justified for small $Fr$ numbers. For all experiments in the $\Gamma=1$ sample, where the $Fr$ number ${Fr} = \Omega^2 (L/2)/g$ is below $0.03$ for all cases, this condition is fulfilled. For the experiments in the $\Gamma=2$ sample the Froude number is below $0.05$ for all experiments up to $1/Ro=5$. For higher $1/Ro$ the Froude number quickly increases up to $0.12$ for $Ra=2.91\times10^8$ and up to $0.49$ for $Ra=1.16\times10^9$. For the experiments at $Ra=2.91\times10^8$, for which the highest $Fr$ number is $0.12$, this does not influence the results as we find a perfect agreement between the experimental and numerical results, see figure 2c. The experimental results for higher $Ra$ in the $\Gamma=2$ sample, where the Froude numbers are higher for the highest $1/Ro$ case, do not suggest a strong influence of this effect on the heat transport measurements. However, at the moment we can not rule it out completely either.

We note that the simulations presented here are very CPU time intensive; about $1.5$ million standard DEISA CPU hours have been used, due to the large aspect ratios, the relatively high $Ra$ numbers that are resolved with high resolution, the number of different $Ro$ number cases, and the long averaging times that are needed to get sufficient statistical convergence.

\section{Results} \label{Sec4}

\begin{figure}
\centering
\subfigure{\includegraphics[width=0.49\textwidth]{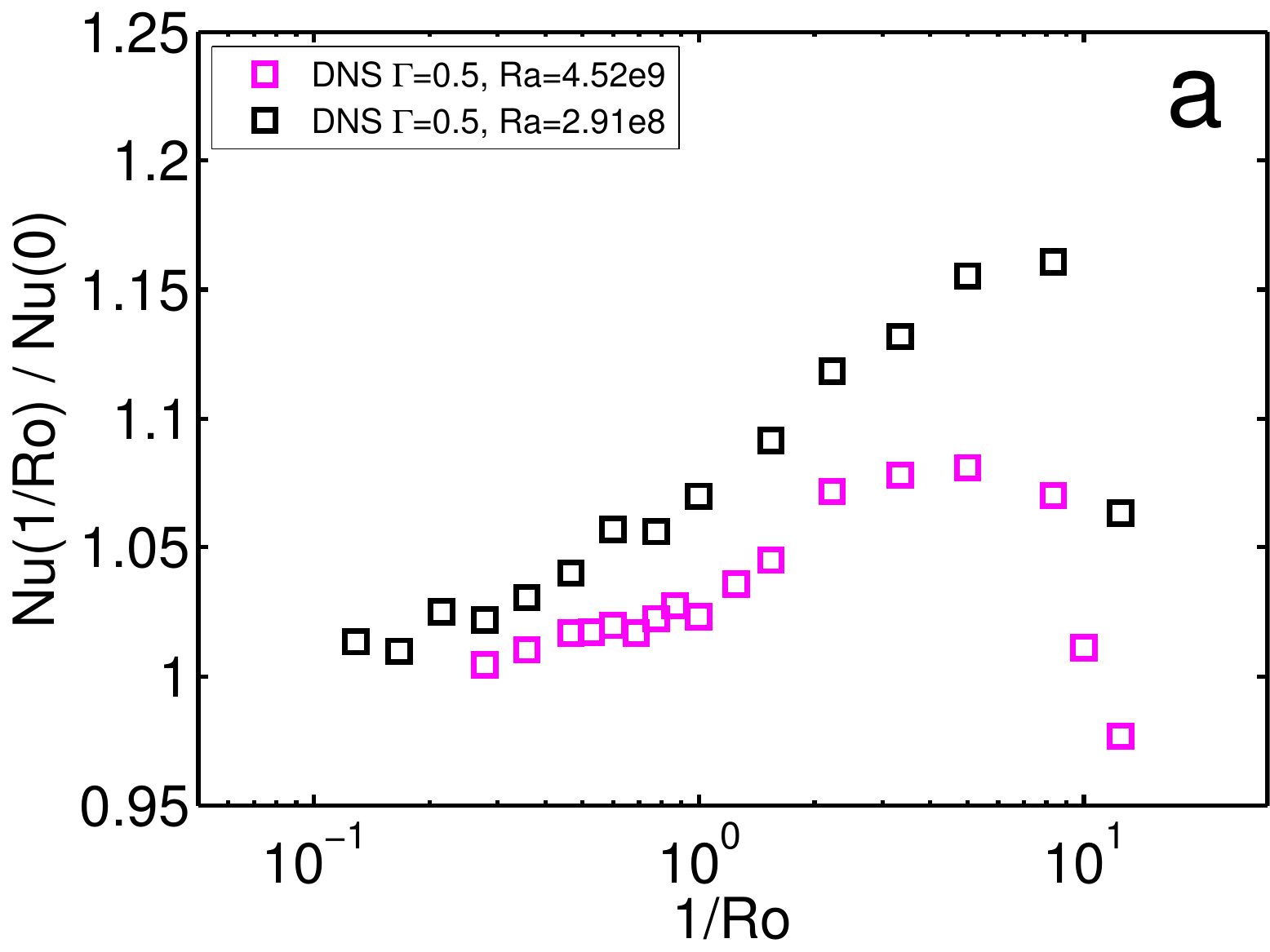}}
\subfigure{\includegraphics[width=0.49\textwidth]{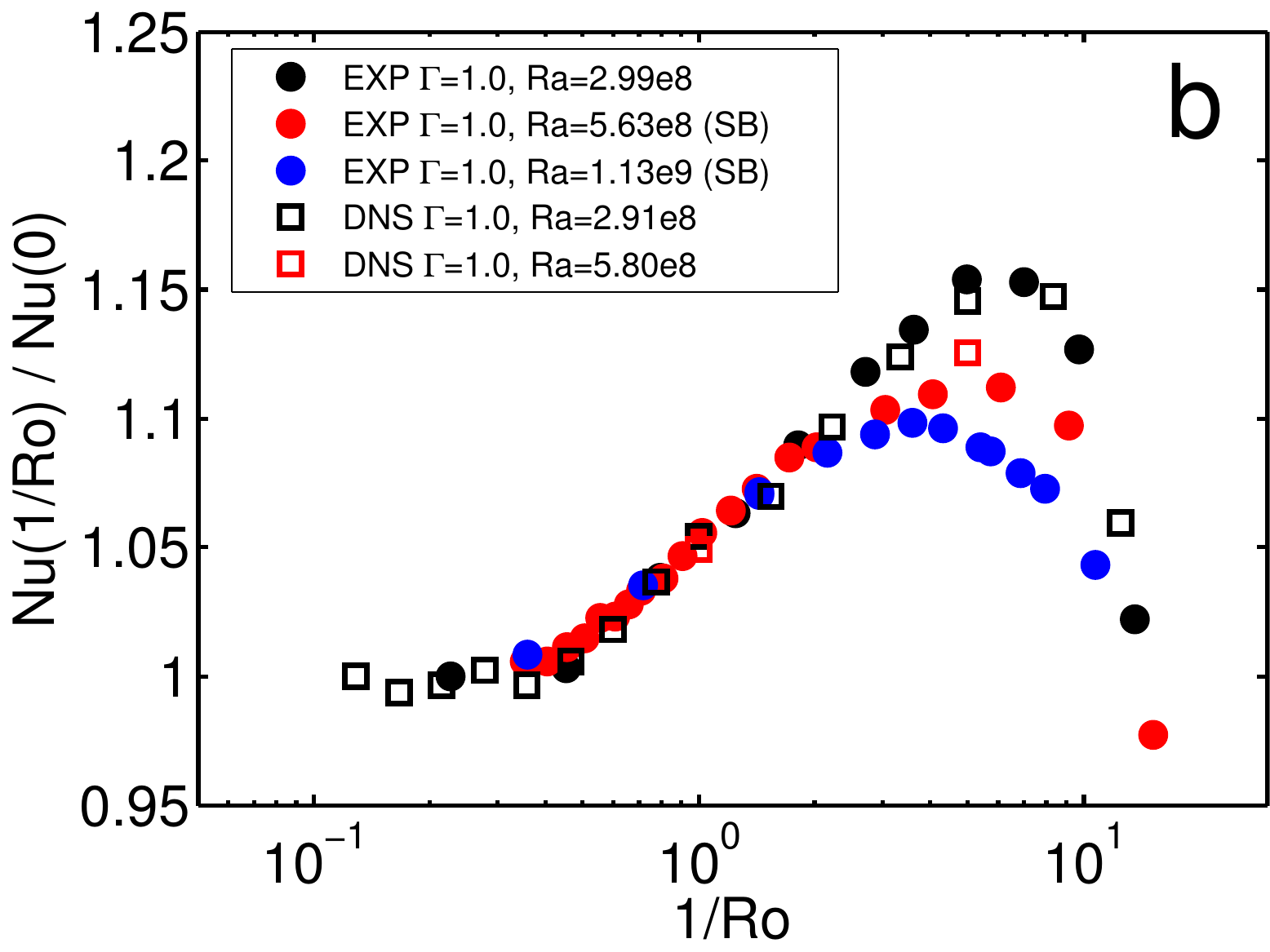}}
\subfigure{\includegraphics[width=0.49\textwidth]{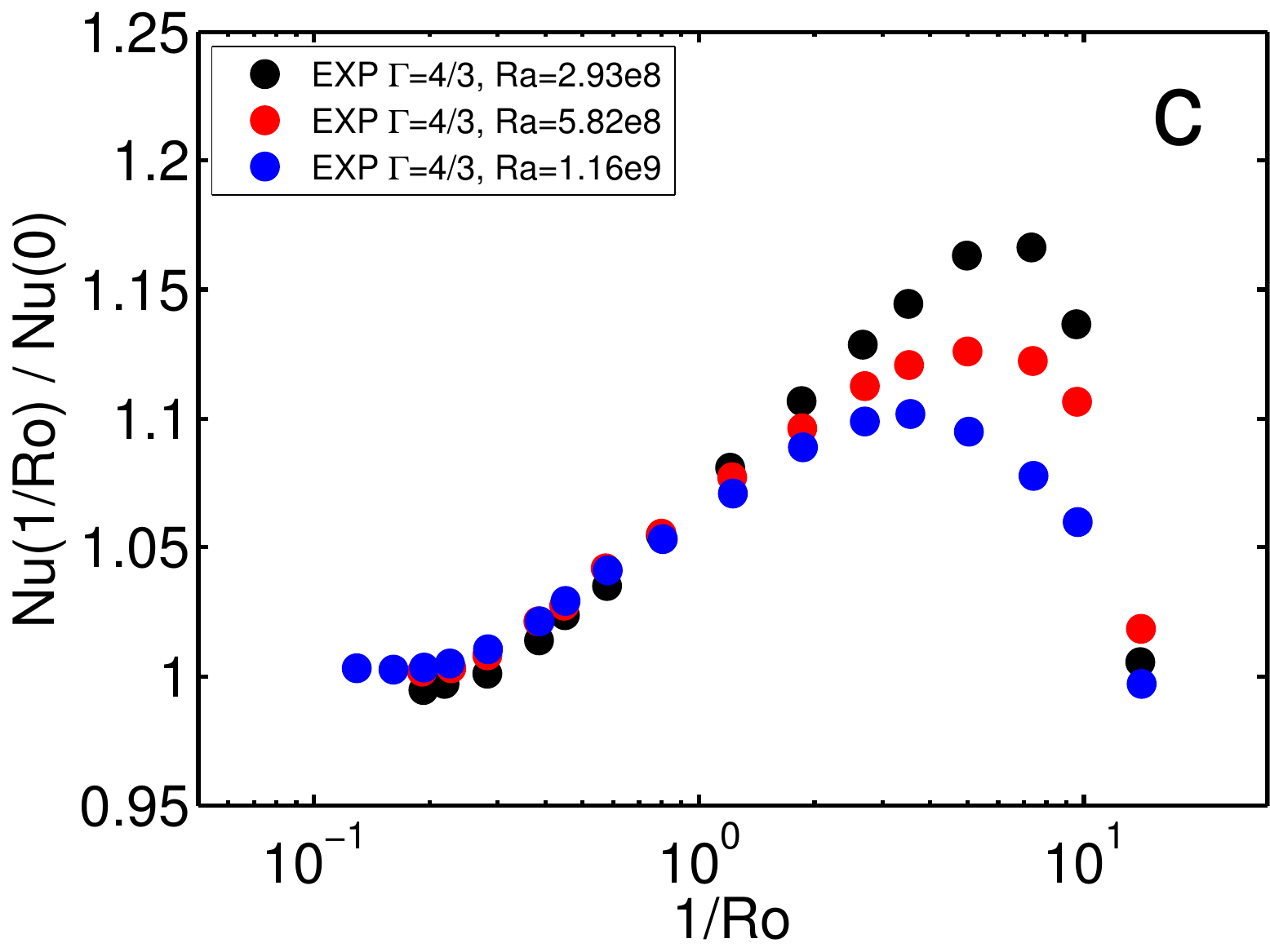}}
\subfigure{\includegraphics[width=0.49\textwidth]{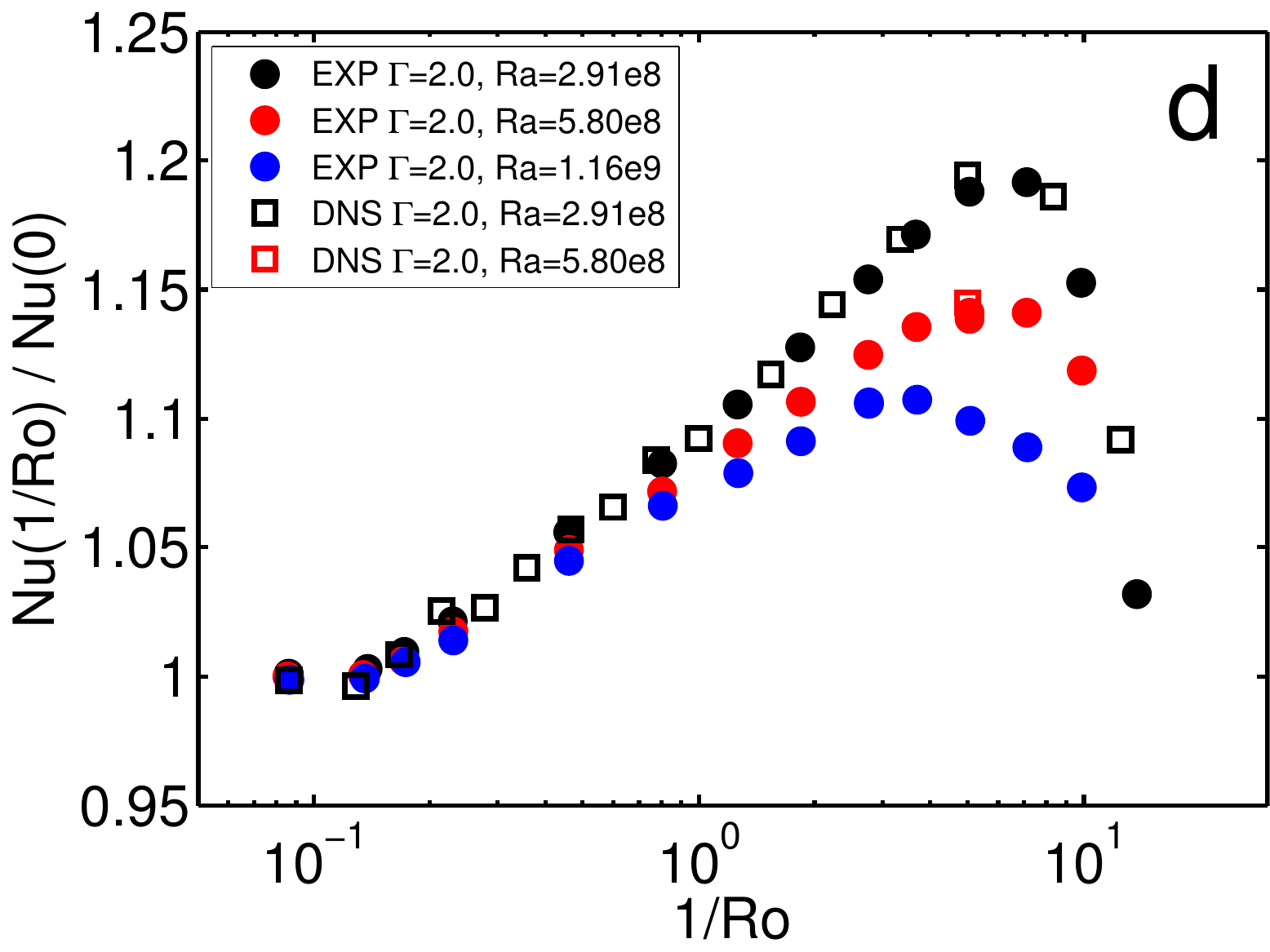}}
\caption{The ratio $Nu(1/Ro)/Nu(0)$ as function of $1/Ro$ for different $Ra$ for a) $\Gamma=0.5,\ $b) $\Gamma=1$, c) $\Gamma=4/3$, and d) $\Gamma=2$.
a) The results from the DNS at $Ra=2.91 \times10^8$, and $Ra=4.52\times10^9$ are indicated by the black and magenta open squares, respectively.
b) The experimental results for $Ra=2.99\times10^8$, $Ra=5.63\times10^8$ (run E4 of Zhong $\&$ Ahlers \cite{zho10c}), and $Ra=1.13\times10^9$ (run E5 of Zhong $\&$ Ahlers \cite{zho10c}) are indicated in black, red, and blue solid circles, respectively. The DNS results for $Ra=2.91 \times10^8$, and $Ra=5.80\times10^8$ are indicated by black and red open squares, respectively.
c) The experimental results for $Ra=2.93\times10^8$, $Ra=5.82\times10^8$, and $Ra=1.16\times10^9$ are indicated in black, red, and blue solid circles, respectively. 
d) The experimental results for $Ra=2.91\times10^8$, $Ra=5.80\times10^8$, and $Ra=1.16\times10^9$ are indicated in black, red, and blue solid circles, respectively. The DNS results for $Ra=2.91 \times10^8$, and $Ra=5.80\times10^8$ are indicated by black and red open squares, respectively.
All presented data in this figure are for $Pr=4.38$.
}
\label{fig:Nusselt_G1_G2}
\end{figure}

\begin{figure}
\centering
\subfigure{\includegraphics[width=0.49\textwidth]{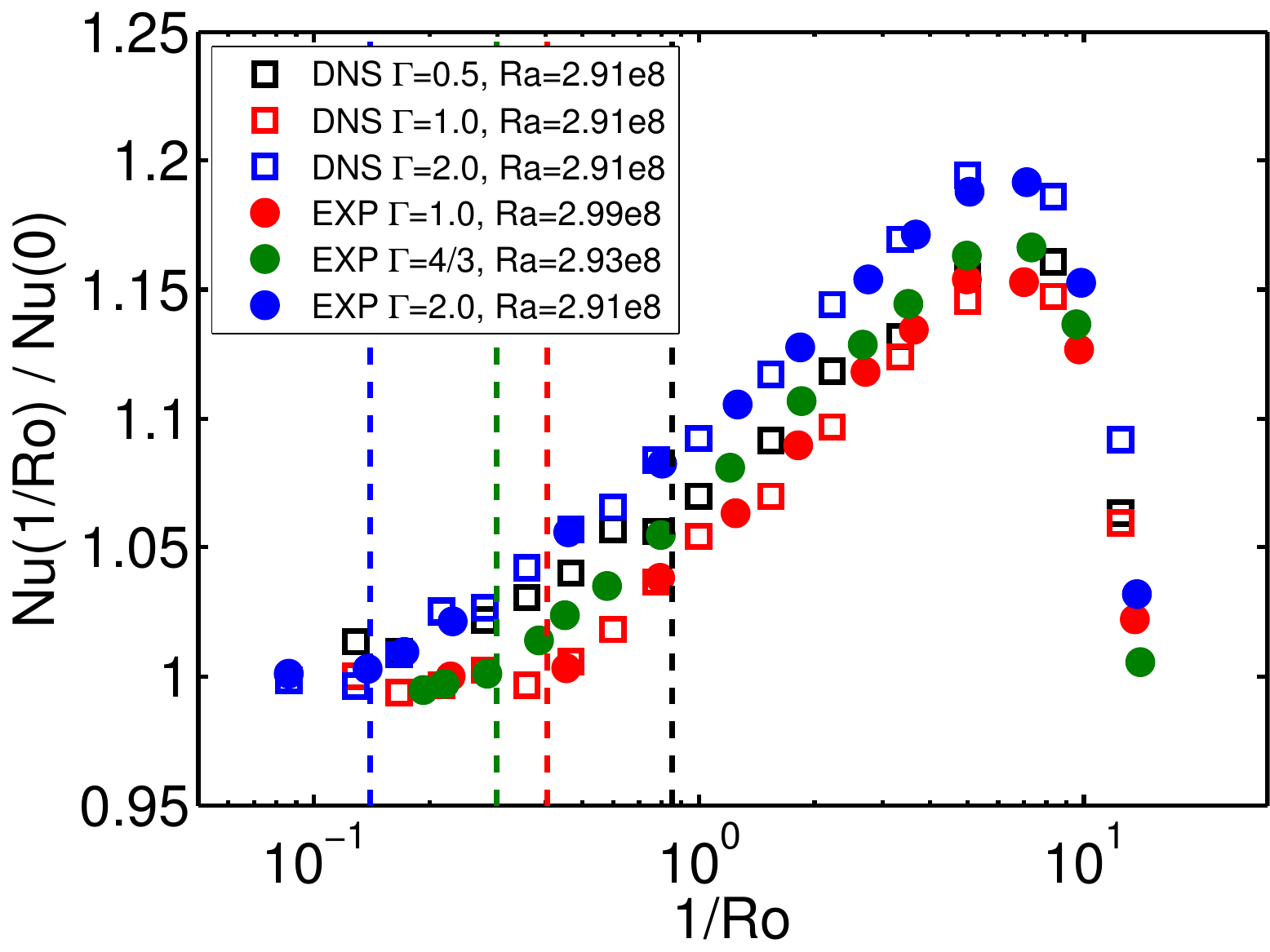}}
\caption{The ratio $Nu(1/Ro)/Nu(0)$ as function of $1/Ro$ for $Ra\approx3\times10^8$ and different $\Gamma$. The experimental results for $\Gamma=1$, $\Gamma=4/3$, and $\Gamma=2$ are indicated in red, dark green, and blue solid circles, respectively. The DNS results for $\Gamma=0.5$, $\Gamma=1$, and $\Gamma=2$ are indicated by black, red and blue open squares, respectively. All presented data are for $Pr=4.38$.}
\label{fig:Nusselt_Ra3e8}
\end{figure}

\begin{figure}
\centering
\subfigure{\includegraphics[width=0.49\textwidth]{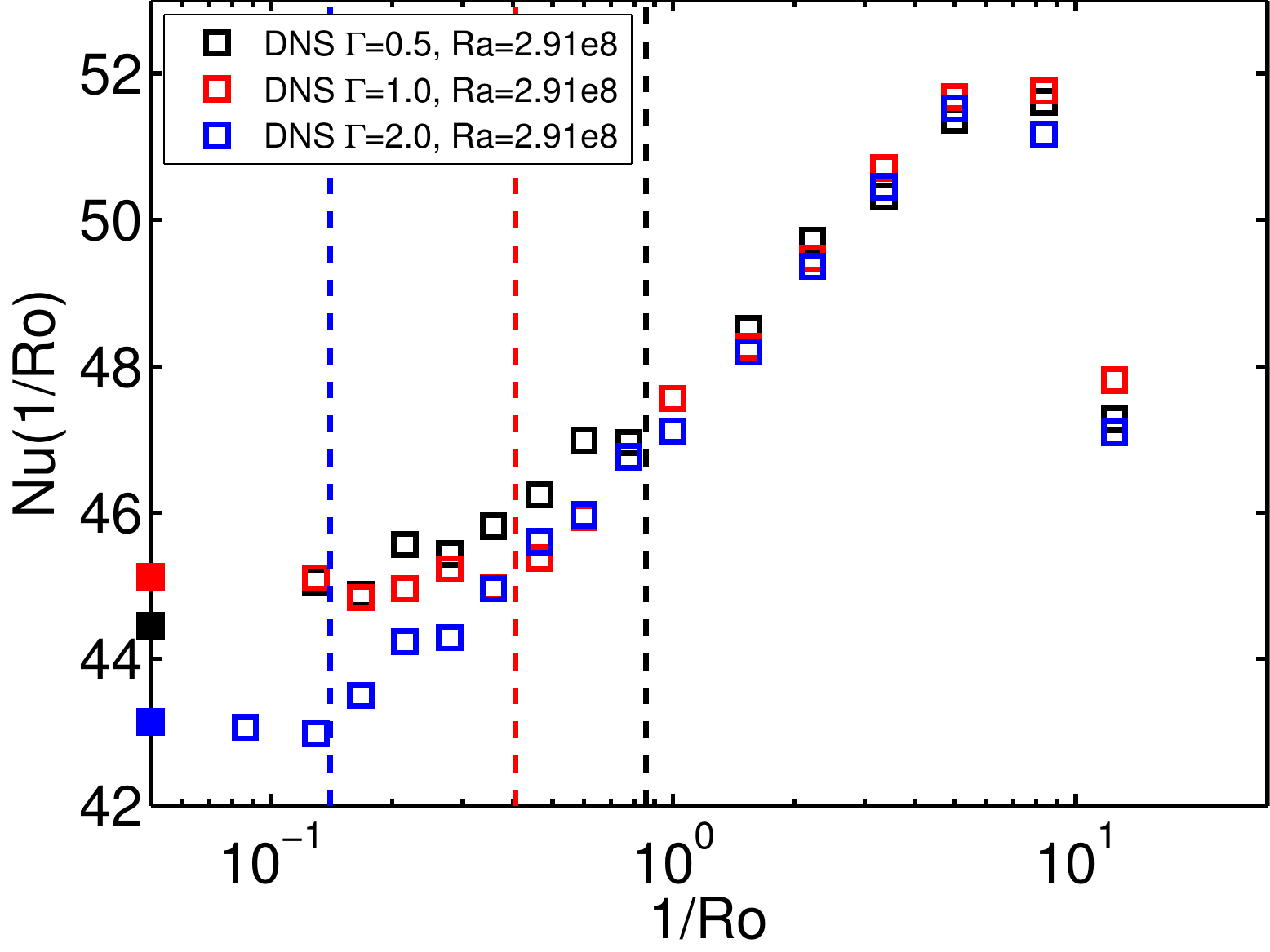}}
\caption{The absolute Nusselt number of the simulation data presented in figure \ref{fig:Nusselt_Ra3e8}. The non-rotating values are indicated by the filled symbols on the left-hand side. The results for $\Gamma=0.5$, $\Gamma=1$, and $\Gamma=2$ are indicated by black, red and blue open squares, respectively. The black, red and blue dashed vertical lines indicate $1/Ro_c$ for $\Gamma=0.5$, $\Gamma=1$, and $\Gamma=2$, respectively. All presented data are for $Pr=4.38$.}
\label{fig:Nusselt_Ra3e8_abs}
\end{figure}

\begin{figure}
\centering
\subfigure{\includegraphics[height=0.49\textwidth]{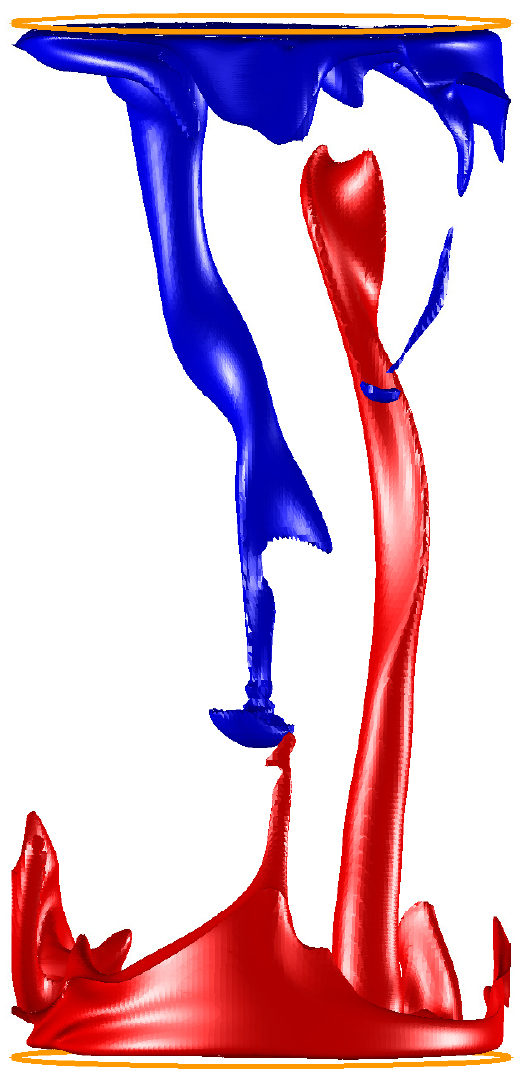}}
\hspace{1.5cm}
\subfigure{\includegraphics[height=0.49\textwidth]{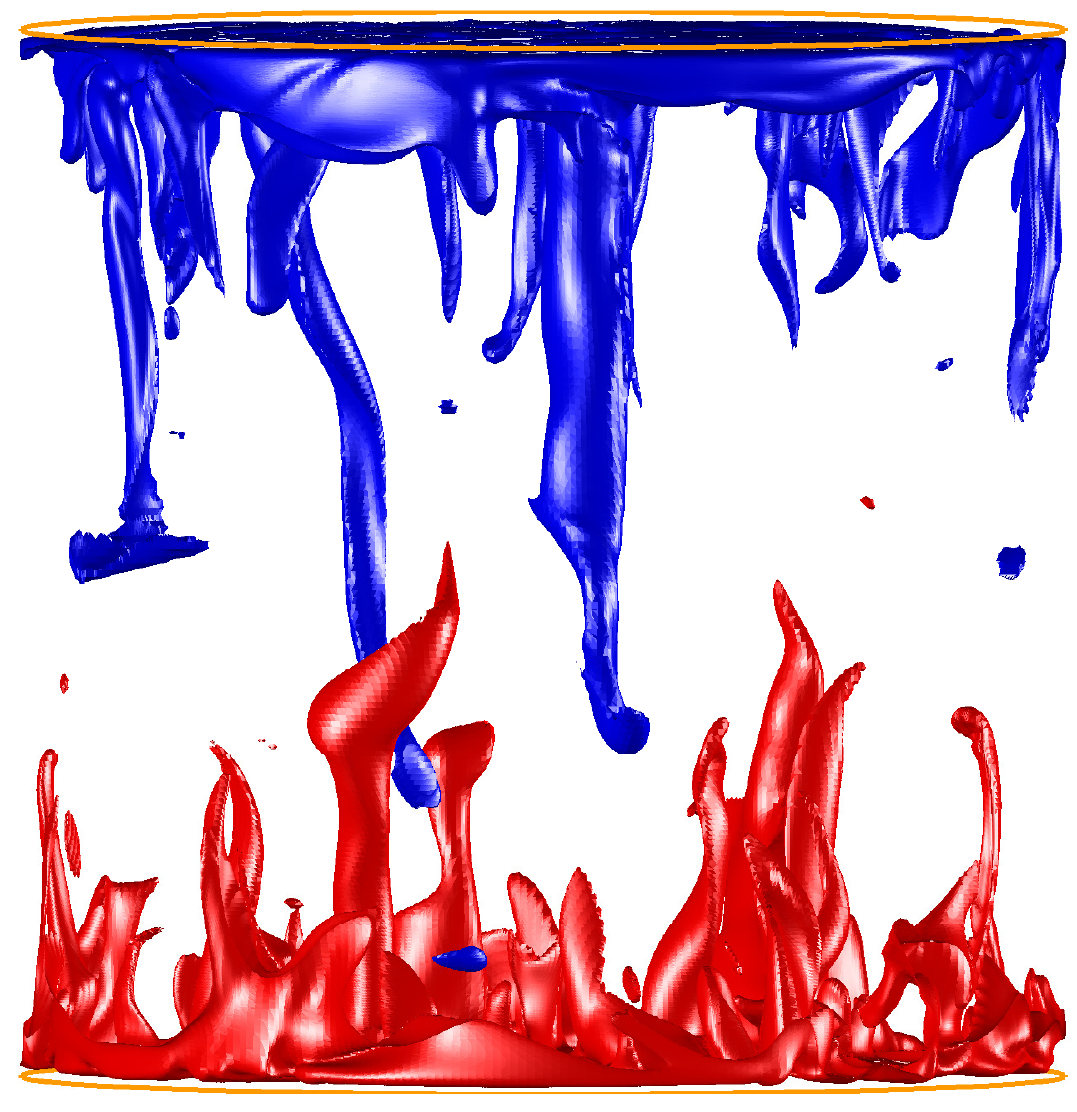}}
\subfigure{\includegraphics[height=0.49\textwidth]{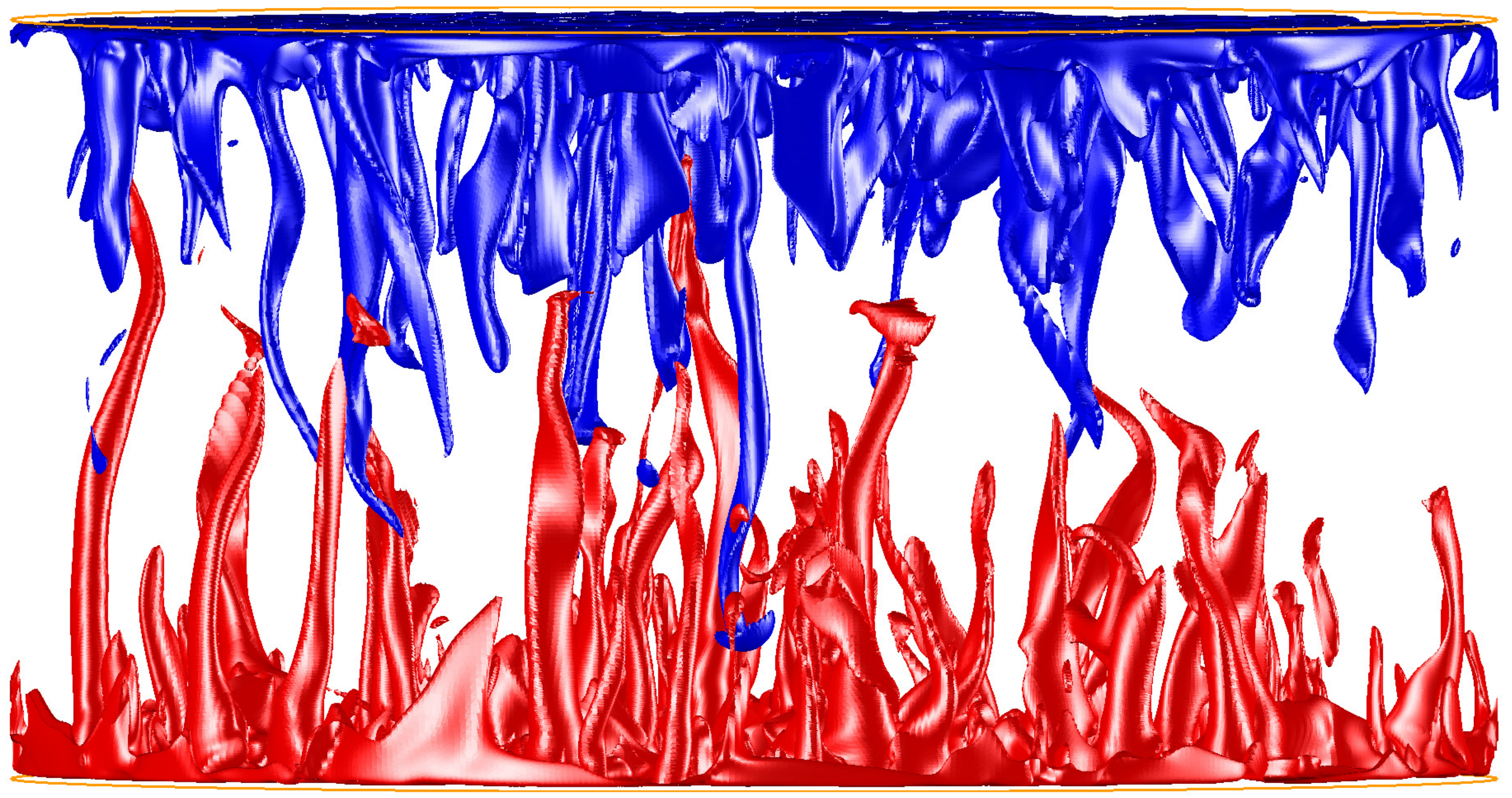}}
\caption{3D visualization of the temperature isosurfaces in the cylindrical sample at $0.65 \Delta$ (red) and $0.35\Delta$ (blue), respectively for $Ra=2.91\times10^8$, $Pr=4.38$, $1/Ro=3.33$ and $\Gamma=0.5$ (left upper plot), $\Gamma=1.0$ (right upper plot), and $\Gamma=2.0$ (lower plot).}
\label{fig:3dpictures}
\end{figure}

\begin{figure}
\centering
\subfigure{\includegraphics[width=0.49\textwidth]{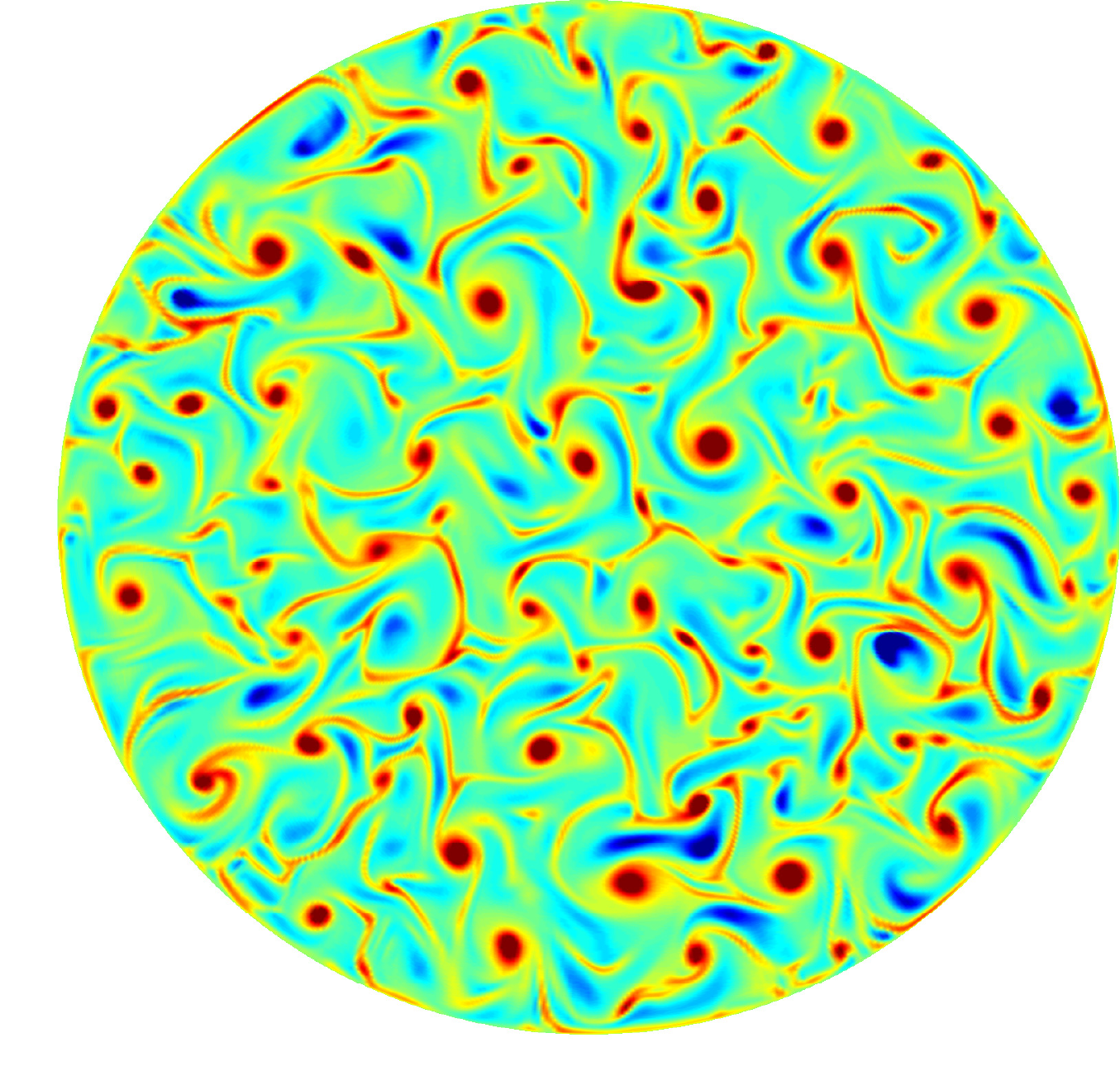}}
\subfigure{\includegraphics[width=0.49\textwidth]{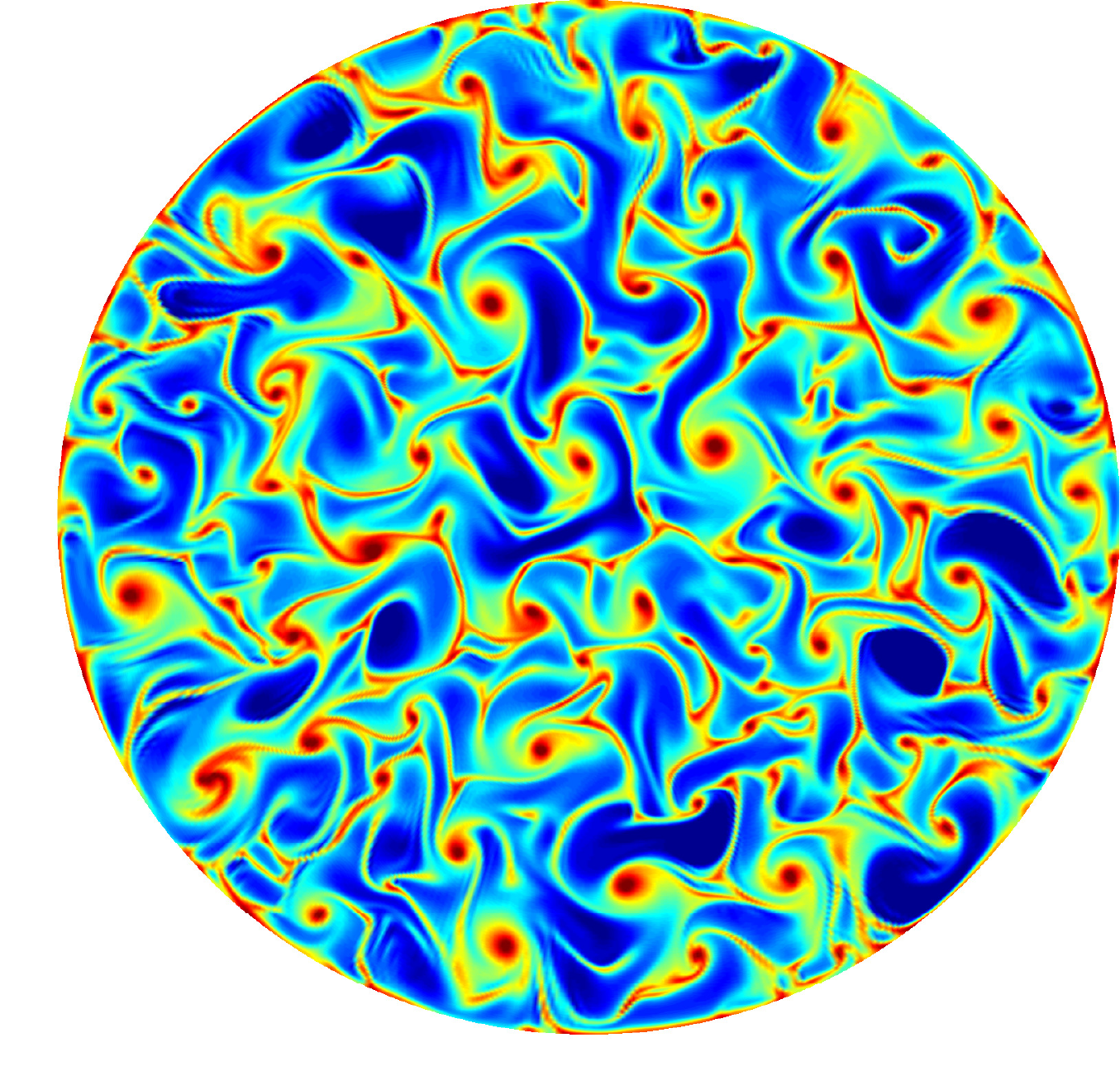}}
\caption{Visualization of the vertical velocity (left plot) and temperature (right plot) fields at the kinetic BL height for $Ra=2.91\times10^8$, $\Pr=4.38$, $1/Ro=5$, and $\Gamma=2$. Note that there is a strong correlation between the areas where the strongest vertical velocity and highest temperatures are found, namely in the vortices. Red and blue indicate upflowing (warm) and downflowing (cold) fluid in panel a (b).
} 
\label{fig:q3_t_fields}
\end{figure}

\begin{figure}
\centering
\subfigure{\includegraphics[width =0.80\textwidth]{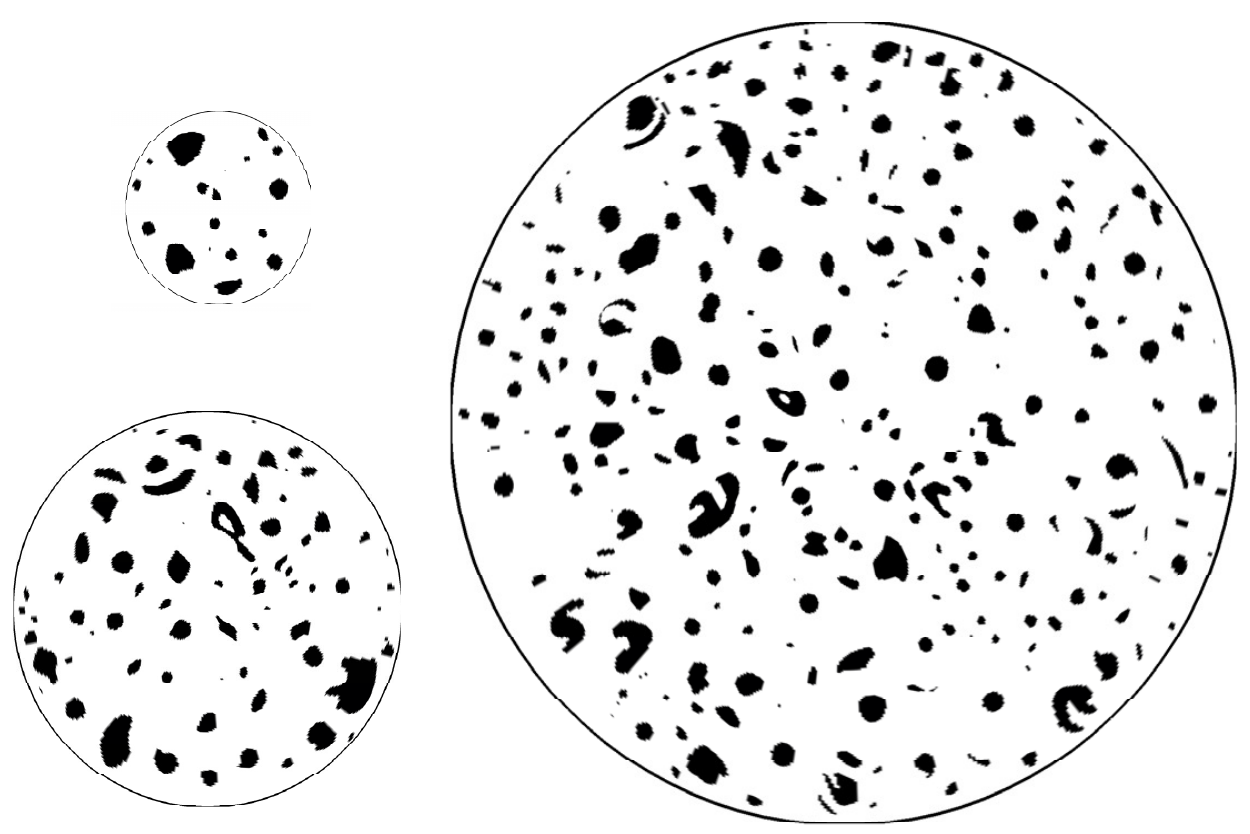}}
\caption{The vortices at the edge of the kinetic BL as identified by the $Q_{2D}$ criterion for $Ra=2.91\times10^8$, $\Pr=4.38$, $1/Ro=5$ and $\Gamma=0.5$ (upper left plot), $\Gamma=1.0$ (lower left plot), and $\Gamma=2.0$ (right plot).}
\label{fig:2dvortex}
\end{figure}

\begin{figure}
\centering
\subfigure[]{\includegraphics[width =0.49\textwidth]{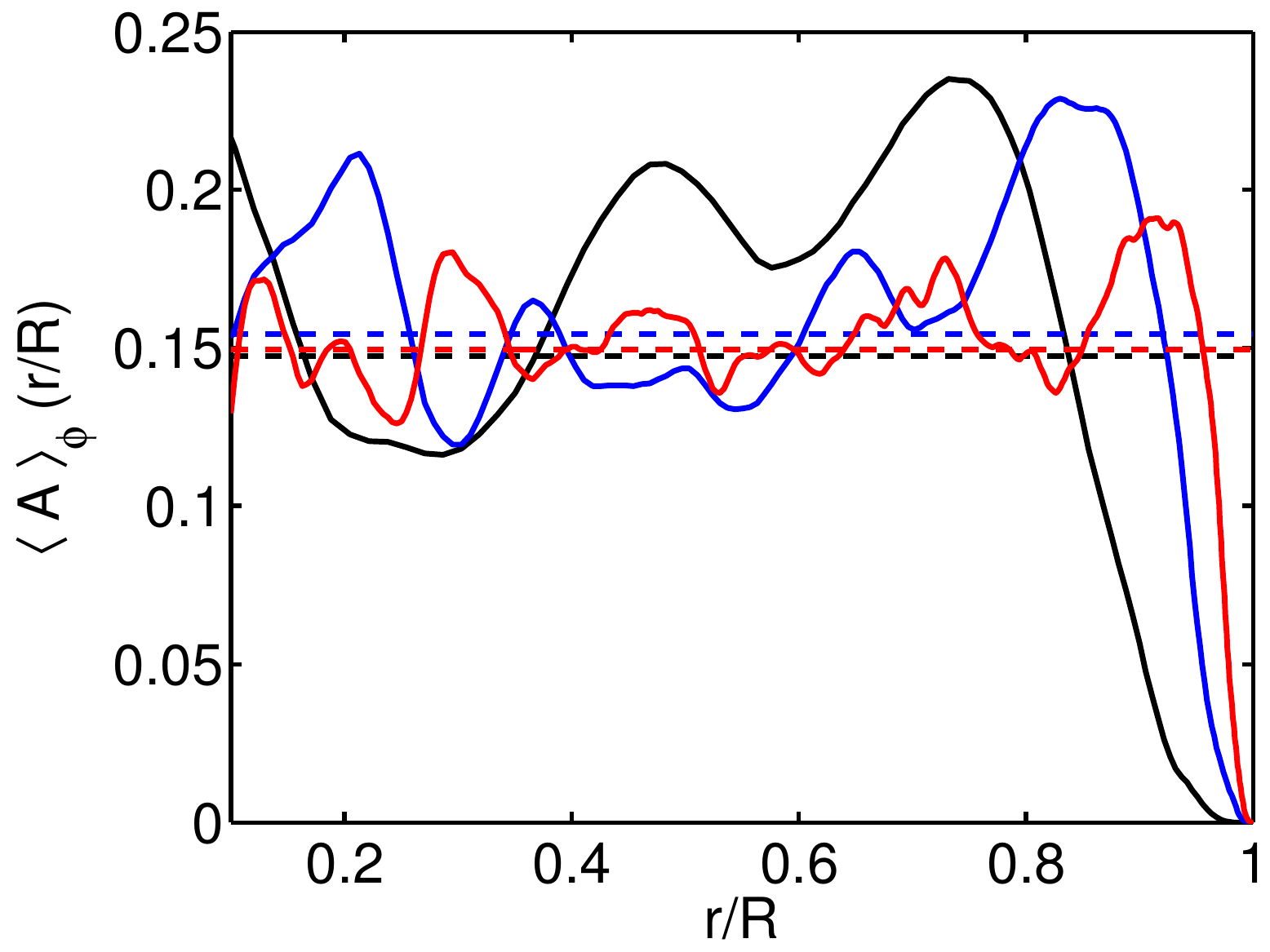}}
\subfigure[]{\includegraphics[width =0.49\textwidth]{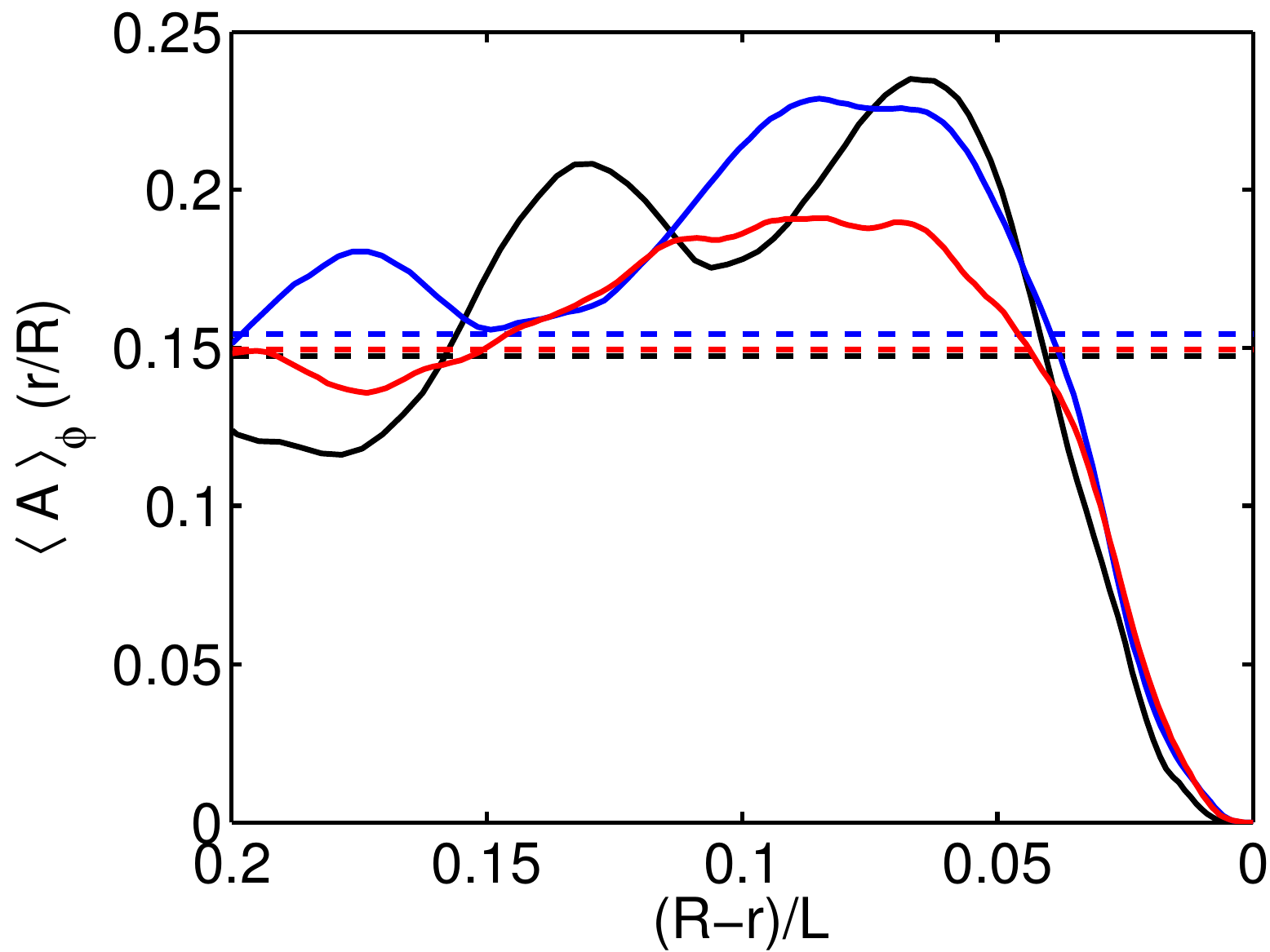}}
\caption{The dashed lines indicate the average of the fraction of the horizontal area A covered by vortices for $Ra = 2.91 \times10^8$, $Pr = 4.38$ and $5 < 1/Ro < 8.33$ and different aspect ratios. Black, blue, and red indicate the data for $\Gamma=0.5$, $\Gamma=1.0$, and $\Gamma=2.0$, respectively. The azimuthal average of the fraction of the area that is covered by vortices as a function of the radial position, i.e. $\langle A \rangle_{r/R}$, is indicated by the solid lines. Panel (a) shows the data as function of $r/R$ and panel (b) as function of $(R-r)/L$ close to the sidewall. For each aspect ratio the data are averaged for the simulation data at $1/Ro=5$ and $1/Ro=8.33$. For each rotation rate $2$ snapshots at the bottom and top BL are used. }
\label{fig:pdf_vortexdistribution}
\end{figure}

In figure \ref{fig:Nusselt_G1_G2} the heat transport enhancement with respect to the non-rotating case is shown for $\Gamma=0.5$, $\Gamma=1$, $\Gamma=4/3$, and $\Gamma=2$ and different $Ra$. Here we find an excellent agreement between experimental and numerical results. The figure shows that there is a strong heat transport enhancement due to Ekman pumping when $1/Ro > 1/Ro_c$, where $1/Ro_c$ indicates the position of the onset of heat transport enhancement \cite{ste09,wei10}. The figure also shows that the heat transport enhancement decreases with increasing $Ra$. This is because the eddy thermal diffusivity is larger at higher $Ra$ and this causes the warm (cold) fluid that enters the base of the vortices to spread out more quickly in the middle of the sample. This makes the effect of Ekman pumping smaller and this results in a lower heat transport enhancement. The breakdown of Nu at high $1/Ro$ is an effect of the suppression of vertical velocity fluctuations through the strong rotation.

When the results for $\Gamma=0.5$, $\Gamma=1$, $\Gamma=4/3$ and $\Gamma=2$ are compared, they look pretty similar at first sight, i.e. a strong heat transport enhancement when $1/Ro > 1/Ro_c$ and a heat transport reduction for very strong rotation rates. However, when we look more closely we see that there are several important differences. First of all $1/Ro_c$, the rotation rate needed to get heat transport enhancement, increases with decreasing aspect ratio. Weiss et al.\ \cite{wei10,wei11} already observed this and explained it by finite size effects. In addition, a close comparison between the $\Gamma=1$ and $\Gamma=2$ data in figure \ref{fig:Nusselt_G1_G2} shows that the heat transport enhancement is slightly larger in a $\Gamma=2$ sample than in a $\Gamma=1$ sample. This can be observed in more detail in figure \ref{fig:Nusselt_Ra3e8}, where the relative 
heat transport enhancement at fixed $Ra$ is compared for several aspect ratios.

The absolute heat transfer shown in figure \ref{fig:Nusselt_Ra3e8_abs} is available only for the numerical case, (as explained in section 2).
In that figure we see that there are some visible differences in the heat transport for the non-rotating case, i.e. without rotation we find that the heat transport in a $\Gamma=2.0$ sample is approximately $5\%$ lower than in a $\Gamma=1.0$ sample. Similar differences in the heat transport as function of $\Gamma$ have been shown by the numerical study of  Bailon-Cuba et al.\ \cite{bai10} and in experiments of Funfschilling et al.\ \cite{fun05} and Sun et al.\ \cite{sun05}, although these experimental and numerical results seem to suggest that the heat transport becomes less dependent on the aspect ratio for higher $Ra$. The main point indicated by figure \ref{fig:Nusselt_Ra3e8_abs} is that the heat transport becomes independent of the aspect ratio once $1/Ro \gg 1/Ro_c$, $1/Ro_c\approx$ $0.14$, $0.4$ and $0.86$ for $\Gamma = 2$, $1$ and $0.5$, respectively. We note that we also find for $Ra=5.80\times10^8$ that the difference in Nusselt between the $\Gamma=1$ and the $\Gamma=2$ case is smaller for $1/Ro=5$ than for the non-rotating case, which is in agreement with the data presented in figure \ref{fig:Nusselt_Ra3e8_abs}.

We believe that the reason for this phenomenon lies in the flow structures that are formed. For the non-rotating case the flow organizes globally in the large scale convection roll. Because this global flow structure can depend on the aspect ratio, there can be small variations in the Nusselt number as function of the aspect ratio. For strong enough rotation, i.e. $1/Ro \gg 1/Ro_c$, the global LSC is replaced by vertically-aligned vortices as the dominant feature of the flow \cite{kun06,kun10,zho09b,ste09,ste10a,zho10c}. In this regime most of the heat transport takes place in vertically-aligned vortices \cite{bou86,sak97,por08,gro10}. Because the vortices are a local effect the influence of the aspect ratio on the heat transport in the system should be negligible. This assumption is used in several models \cite{leg01,por08,gro10}, which consider a horizontally periodic domain, that are developed to understand the heat transport in rotating turbulent convection.

To investigate this idea we made three-dimensional visualizations of the temperature isosurfaces at $Ra=2.91\times10^8$, $Pr=4.38$, and $1/Ro=3.33$, for the different aspect  ratios, see figure \ref{fig:3dpictures}. Indeed the figures confirm that vertically-aligned vortices are formed in all aspect-ratio samples. Furthermore, the figure reveals that a larger number of vortices is formed in the $\Gamma=1$ and $\Gamma=2$ samples than in the $\Gamma=0.5$ sample. This is expected, since these samples have a larger horizontal extension.

In order to reveal the structure of the flow in more detail we show the temperature and vertical velocity fields at the kinetic BL height near the bottom plate for $Ra=2.91\times10^8$, $Pr=4.38$, $1/Ro=5$, and $\Gamma=2$ in figure \ref{fig:q3_t_fields} . This figure clearly shows that hot fluid is captured in the up-going vortices. Similar plots (not shown) revealed that cold fluid near the top plate is captured in down going vortices. In order to study the vortex statistics one needs to have a clear criterion of what exactly constitutes a vortex. For this we use the so-called $Q$-criterion \cite{bou86,vor98,vor02,kun10b,wei10}. This criterion requires that the quantity $Q_{2D}$ \cite{mcw84,ste09}, which is a quadratic form of various velocity gradients, is calculated in a plane of fixed height. Here we always take the kinetic BL height. Following Weiss et al.\ \cite{wei10} an area is identified as "vortex'' when $Q_{2D}<-\langle |Q_{2D}|\rangle_{v}$, where $\langle |Q_{2D}|\rangle_{v}$ is the volume-averaged value of the absolute values of $Q_{2D}$. Here we have set the threshold for the vortex detection more restrictive than in \cite{wei10} to make sure that only the strong upgoing (downgoing) vortices are detected. However, we note that similar results are obtained when a less restrictive threshold is used.

The result of this procedure for $Ra=2.91\times10^8$, $1/Ro=5$, and $\Gamma=0.5-2.0$ is shown in figure \ref{fig:2dvortex}. The figure shows that the vortices (both up and down going) are in general randomly distributed. However, note that no vortices are formed close to the sidewall. To quantify this we determined the radial distribution of the vortices from the plots shown in figure \ref{fig:2dvortex} and similar plots. The result is shown in figure \ref{fig:pdf_vortexdistribution}. Figure \ref{fig:pdf_vortexdistribution}a confirms that no vortices are formed close to the sidewall, while in the bulk their fraction is roughly constant. Figure \ref{fig:pdf_vortexdistribution}b shows that the size of the region close to the sidewall where no vortices are formed is roughly independent of the aspect ratio.
This is in agreement with the predication of Weiss $\&$ Ahlers \cite{wei11} that is derived from a phenomenological  Ginzburg-Landau model. As detailed information about the flow field is needed to determine the vortex distribution it is very hard to obtain this data from experimental measurements \cite{kun10b}. In figure \ref{fig:pdf_vortexdistribution}b one can see that in the region $(R-r)/L\lesssim 0.015$ the value of $\langle A \rangle_{r/R}$ decreases faster to zero than for $(R - r)/L \gtrsim 0.015$, which is due to the vortex detection method employed in this study. More specifically, we detect only the core of the vortex. As the vortex core is always formed some distance away from the wall this causes an (artificial) enhanced decrease in the number of vortices that are detected in the direct vicinity of the wall. Due to the vortex detection method that is used it is also difficult to estimate the average radius of the vortices as only the core of the vortex is detected.

However, and this is the main point here, one can see in figure \ref{fig:pdf_vortexdistribution} that the fraction of the horizontal area that is covered by vortices, see the dashed lines in the figure, is independent of the aspect ratio. This observation supports our finding that the heat transport is independent of the aspect ratio in the rotating regime. This result may seem somewhat unexpected based on the data shown in figure \ref{fig:pdf_vortexdistribution}b. This figure namely shows that the absolute size of the vortex depleted region close to the sidewall is approximately aspect ratio independent. Therefore one would have expected that the average horizontal area that is covered by vortices averaged over the whole area is higher for larger aspect ratio, because for larger aspect ratio samples this vortex-depleted sidewall region is relatively smaller than for smaller aspect ratio ones. However, just next to the vortex depletion region we find the vortex enhanced region. As is shown in figure \ref{fig:pdf_vortexdistribution}b, the absolute width of this region seems to be rather independent of the aspect ratio of the sample, too. Hence the effect of the vortex depletion and the vortex enhancement region on the horizontally averaged area that is covered by vortices cancel out in first order. 

In this paper we focused on the the influence of the aspect ratio on the fraction of the horizontal area that is covered with vortices. We note that Weiss et al.\ \cite{wei10} have already investigated the effect of rotation rate on fraction of the horizontal area that is covered by vortices. There we showed that the average horizontal area that is covered by vortices increases approximately linear with $1/Ro$ when $1/Ro>1/Ro_c$. This result is in agreement with the predictions obtained from a phenomenological Ginzburg-Landau like model that is discussed in that paper. Furthermore, it was shown by Kunnen et al.\ \cite{kun10b} that the vortex densities and mean vortex radii are mostly independent of the Taylor number $Ta=Ra/(Ro^2Pr)$ except very close to the bottom and top plates where more vortices are detected when the Taylor number is raised. As the numerical simulations considered here have been obtained for similar $Ra$ and $Pr$ as the ones used in the above studies we have not investigated the influence of these parameters on the vortex statistics.

\section{Discussion}
In summary, we investigate the effect of the aspect ratio on the heat transport in turbulent rotating Rayleigh-B\'enard convection by results obtained from experiments and direct numerical simulations. We find that the heat transport in the rotating regime is independent of the aspect ratio, although there are some visible differences in the heat transport for the different aspect ratios in the non-rotating regime at $Ra=2.91 \times 10^8$. This is because in the non-rotating regime the aspect ratio can influence the global flow structure. However, in the rotating regime most heat transport takes place in vertically-aligned vortices, which are a local effect. Based on the simulation results we find that the fraction of the horizontal area that is covered by the vortices is independent of the aspect ratio, confirms that the vertically-aligned vortices are indeed a local effect. This supports the simulation results, which show that the heat transport becomes independent of the aspect ratio in the rotating regime. In addition, it confirms the main assumption that is used in most models, which consider a horizontally periodic domain \cite{leg01,por08,gro10}, that are developed to understand the heat transport in rotating turbulent convection. The analysis of the vortex statistics also revealed that the  vortex concentration is reduced close to the sidewall, while the distribution is nearly uniform in the center. In between these two regions, there is a region of enhanced vortex concentration. The widths of both that region and the vortex-depleted region close to the sidewall are  independent of the aspect ratio. This analysis highlights the value of numerical simulations in turbulence research: The determination of the vortex distribution requires detailed knowledge of  the flow field and therefore  it would have been very difficult to obtain this finding purely from experimental measurements.

\begin{acknowledgments}
\noindent
{\it Acknowledgment:} We gratefully acknowledge various discussions with Guenter Ahlers over this line of research and his helpful comments on our manuscript. The authors wish to thank Eric de Cocq, Gerald Oerlemans, and Freek van Uittert (design and manufacturing of the experimental set-up) for their contributions to this work, and
 Jaap van Wensveen of Tempcontrol for advice and helping to calibrate the thermistors. We thank Chao Sun for stimulating discussions.  We thank the DEISA Consortium (www.deisa.eu), co-funded through the EU FP7 project RI-222919, for support within the DEISA Extreme Computing Initiative. We thank Wim Rijks (SARA) and Siew Hoon Leong (Cerlane) (LRZ) for support during the DEISA project. The simulations were performed on the Huygens cluster (SARA) and HLRB-II cluster (LRZ). RJAMS  was financially supported by the  Foundation for Fundamental Research on Matter (FOM). 
\end{acknowledgments}

\end{document}